\pdfoutput=1

\documentclass[11pt,a4paper]{article}

\usepackage{booktabs}

\usepackage[biblatex]{msws}

\bibliography{msws}

\title{Middle-Square Weyl Sequence RNG}

\author{Bernard Widynski}
\mailto{mswsrng@gmail.com}

\begin{document}

\maketitle

\begin{abstract}
In this article, we propose a new implementation of John von Neumann's middle-square random number generator (RNG).  A Weyl sequence keeps the generator running through a long period.  
\end{abstract}

\section{Introduction}

Early in the field of computer science, John von Neumann proposed the middle-square method for generating random numbers \cite{Neumann}.  A number is squared and the middle digits are returned as the next random number.   This method was used as a source of data in the early work on Monte Carlo. Good data was produced but the middle square had a known problem which was referred to as the \lq \lq zero mechanism\rq \rq.  Once the middle digits became zero the generator would continue to produce zero outputs.  Repeating cycles of non-zero data could also be produced.  In this article we propose the use of a Weyl sequence to run the middle square.  The Weyl sequence overcomes the \lq \lq zero mechanism\rq\rq.  It keeps the generator running through a long period. 

\section{Weyl Sequence}

We use a Weyl sequence similar to the Weyl sequence in Brent's xorgens RNG \cite{Brent}.  This is actually an integer stepping sequence of period $2^{64}$.  In C, it has the following form:   w += s; where w is an unsigned 64-bit integer and s is an odd constant.  A middle-square RNG run by this sequence is shown below:

\begin{verbatim}

uint64_t x = 0, w = 0, s = 0xb5ad4eceda1ce2a9;
inline static uint32_t msws32() {
   x *= x; x += (w += s); return x = (x>>32) | (x<<32);
}

\end{verbatim}

The Weyl sequence is added to the square of x.  The middle is extracted by shifting right 32 bits.  

\section{Middle-Square Implementation}

One might ask \lq \lq Where is the middle?\rq \rq $ $ in the above program.  In C, arithmetic on unsigned 64-bit integers is modulo $ 2^{64}$.  After squaring, the x value will have only the lower 64 bits of the result.  These 64 bits represent the lower half of the square.  Shifting right 32 bits will return the upper part of this lower half.  This is the 32-bit middle that is returned.

Perhaps an illustration might be useful.  We will consider only the square and shift operations.  Let's begin by showing how John von Neumann's original middle square would obtain a middle from a square.  Let's square a 16-digit hexadecimal number, say E3296D171EC4A36F.  The square is \newline

\centerline{C9927E2B2075471D31C2914AAE4E8A21} 
$ $ \newline
The middle is \newline
\centerline{2075471D31C2914A}
$ $ \newline
Our computer, however, is only able to store 64-bit numbers.  That is enough for 16 hexadecimal digits.  So, after squaring x will store only the lower 16 digits of the 32 produced: \newline

\centerline{31C2914AAE4E8A21}
$ $ \newline
The circular shift operation will transform this to \newline

\centerline{AE4E8A2131C2914A}
$ $\newline
The return statement will provide the lower 32 bits or 8 digits:\newline

\centerline{31C2914A}
$ $\newline
Notice that this is the lower half of the middle from the original middle square.  We are in effect returning middle digits from a square. The modulo $ 2^{64}$ arithemetic implicit in C allows us to work with larger squares than would otherwise be possible.  It is not necessary to store all 32 digits of the result.  It is sufficient to use the lower 16 that are computed. This half-square implementation is quite efficient and allows for an easy extraction of the middle by a simple shift operation.  The RNG compiles into only four machine instructions:  imulq, iaddq, iaddq, and rorq.  The rorq is the rotate right instruction which corresponds to a circular shift.  Being so compact, the RNG is suitable for inline which increases efficiency by removing the subroutine linkage.

\section{Overcoming the Zero Mechanism}

The primary defect with the middle square is that it eventually produces repeating cycles of data, often a cycle of zero outputs.  In this section we demonstrate that the Weyl sequence w prevents repeating cycles in x.  We will show this in two steps.  First we demonstrate that no two of the first $ 2^{64}$ elements of the Weyl sequence w can be equal.  It follows from this that there can be no repeating cycles in x. \newline\newline
Theorem A:  For the Weyl sequence that is generated by the formula \mbox {w += s} where w is an unsigned 64-bit integer and s is an odd constant, no two of the first $ 2^{64}$ generated elements can be equal.\newline\newline
Method: This can be shown with a proof by contradiction.  The assumption that any two are equal leads to the contradiction that odd is equal to even. \newline\newline
Proof: The mathematical equivalent of w += s in C is shown below: \newline\newline
For i = 0 to $2^{64}-1$, and some odd constant s $< 2^{64}$ \newline \newline
\indent $ w(i) = (i * s)\; mod\; 2^{64} $\newline \newline
Now, assume that there exists an m and n with m $<$ n $< 2^{64}$ such that\newline \newline
\indent $w(m) = w(n)$  \newline  \newline                           
This implies that\newline \newline
\indent $n * s - m * s = k * 2^{64}$ \hspace{5 mm} for some integer k\newline\newline
which implies that \newline \newline
\indent $(n - m) * s = k * 2^{64}$ \newline \newline 
There are two cases to be considered. \newline \newline
Case 1: (n - m) is odd\newline \newline
Here we have odd times odd equal to some k times even, which implies that odd is equal to even.\newline\newline
Case 2: (n - m) is even\newline \newline
Let $2^{j} * p$  be the factorization of (n - m)  for some integer j and odd integer p.  This will give\newline\newline
\indent $2^{j} * p * s = k * 2^{64}$ \newline\newline 
which is equivalent to \newline\newline
\indent $p * s = k * 2^{64-j}$\newline\newline
Note that $ 0 < j < 64 $ because (n - m) is even and (n - m) is $< 2^{64}$ \newline\newline
which implies that odd equals to even. \newline\newline
Q.E.D.\newline\newline
We conclude that no two of the first $2^{64}$ numbers can be equal.  \newline\newline
From Theorem A it can be determined that the period of the Weyl sequence is $2^{64}$.  If no two of the first $2^{64}$ numbers are equal then all must be different.  The only way this could happen is if each number between 0 and $2^{64}-1$ occurs exactly one time.  Also, from $ w(i) = (i * s)\; mod\; 2^{64} $ we see that the next $2^{64}$ numbers generated will be exactly the same sequence as the first $2^{64}$ numbers generated and for all $2^{64}$ subsequently generated.   It follows that the period of the Weyl sequence is $2^{64}$ and that the output of the Weyl sequence will be uniform.
\newline\newline
Theorem B:  For the first $2^{64}$ iterations of the RNG, there will be no repeating cycles in x.\newline\newline
Method: This can be shown quite easily by noting that the following Weyl sequence value for any two equal x values will be different (Theorem A), thus producing a different x value.  \newline\newline
Proof: Consider two separate x values in the first $2^{64}$ iterations that are equal.  The RNG performs three operations: square, add Weyl sequence, and shift.  The first operation is the square and this will yield the same result.  The next operation is to add the Weyl sequence value w which we know from Theorem A is different.  So the result will be different. The third operation is the circular shift.  The numbers will still be different after shifting. This implies that the next x value after each of the equal x values will be different.  We conclude that there can be no repeating cycles in x for the first $2^{64}$ iterations of the RNG. \newline\newline
Q.E.D.  \newline\newline
That there are no repeating cycles in the the first $2^{64}$ interations implies that the period of x will be at least $2^{64}$.

\section{Uniformity}

In addition to providing a long period, the Weyl sequence also provides a basis for uniformity in the output. The Weyl sequence is uniform.  We demonstrate that by simply adding it to the square we will obtain a uniform output.\newline\newline
Theorem C:  Let x be a random but not necessarily uniform bit stream.  Let w be a uniform but not necessarily random bit stream.  Then x + w will be a uniform bit stream.\newline\newline
Method:  We will work with 32-bit quantities because the output of the RNG is 32 bits.  We can demonstrate uniformity by showing that the probability of any given output is 1/$2^{32}$. \newline\newline
Proof: Given an arbitrary element y in [0, 0xffffffff], let's compute the probability of y where y = x + w. 
$$prob(y) = \sum_{x=0}^{2^{32}-1} (prob(x)*prob(w))$$ \newline
Now because w is uniform, prob(w) will be $1/2^{32}$.  So, 
$$prob(y) = \sum_{x=0}^{2^{32}-1} (prob(x)*1/2^{32})$$ \newline
which is equvalent to
$$prob(y) = 1/2^{32} \sum_{x=0}^{2^{32}-1} (prob(x))$$ \newline
From basic probability, we know that the sum of the probabilities of all possible outcomes is 1.  So,
$$prob(y) = 1/2^{32}$$
Q.E.D. \newline\newline
There are $2^{32}$ possible outputs and each has the probability of $1/2^{32}$.  We conclude that the RNG will produce a discrete uniform distribution. %\newline\newline

\section{Initialization}

The state vector consists of three 64-bit words: x, w, and s.  Only w and s may be used to create a distinct initialization.  If x is used, there is some danger that overlapping data will be produced.  For example, let's say that one initialized the state vector as follows:  \newline
\begin{center}x = 0; w = 0; s = 0xb5ad4eceda1ce2a9;\end{center} $ $\newline
for the first run and then set the state \newline
\begin{center}x = $2^{32}$; w =0; s = 0xb5ad4eceda1ce2a9;\end{center} $ $ \newline
for the next run.  In this case, exactly the same data would be generated even though the initial x values are different.  0 squared is the same as $2^{32}$ squared when taken mod $2^{64}$. The constant s will provide a distinct initialization with no overlapping data.  Also, w may be used but one must assure sufficient space in between the values so there is no overlap.  Using w for jump ahead will be discussed in a following section.

The constant s should be an irregular bit pattern with roughly half of the bits set to one and half of the bits set to zero.  The best statistics have been obtained with this type of pattern.  Unduly sparse or dense values do not produce good results.  If one initializes the RNG, for example, as follows: \mbox{x=0; w=0; s = 0x0000000100000001;} the first set of outputs appear non-random: 
\begin{verbatim}
                      00000001
                      00000004
                      0000001b
                      00000406
                      00170a61
                      f765b52a
                      68d57352
                      0aafc03f
                      f461cd1e
                      fbe33cc0
                      808d47e0
                      230dc324
                      93202f86
                     \end{verbatim} 

The numbers eventually square out to random data, but it takes about five iterations before this happens.   This type of effect can be avoided by choosing s constants that are neither too sparse nor too dense. 

A method that has been shown to work well is to choose the constant s so that the hexadecimal digits or nibbles are different.  This results in change in each 4-bit section of the Weyl sequence on each iteration of the RNG.  The following constants are of this type: 0x9f32e1cbc5e1374b, 0x278c5a4d8419fe6b, 0x38ea2514b48de29f, 0x91c43526df517a8b. The digits are chosen so that the upper 8 are different and the lower 8 are different.  Only non-zero digits are used and the least significant digit is chosen to be odd. Such constants have been tested extensively and produce good data.   A utility called init\_rand\_digits which creates this type of constant is provided in the software download\footnote{Software download available at http://mswsrng.wixsite.com/rand}.  There are two methods for use.  One may simply call init\_rand\_digits from the program to create the constant.  Or one may use init\_rand\_digits offline to create an include file of constants that could be compiled into the program. The compiled in constants would make for a faster initialization.  This routine has been verified to produce unique constants for input values ranging from 0 to 3 billion.  Example usage is shown below: \newline

\begin{verbatim}
#include "init.h"
x = w = s = init_rand_digits(n);
\end{verbatim} $ $\newline
or if using pre-computed constants \newline

\begin{verbatim}
uint64_t seed[] = {
#include "seed.h"
};
x = w = s = seed[n];
\end{verbatim}$ $ \newline
Note: x and w have been initialized non-zero so that a randomization occurs on the first iteration.  This assures that as a set the first outputs from all streams should be random data.

The number of available constants from init\_rand\_digits is on the order of 33 quadrillion.  Since the stream length is $2^{64}$, this provides about $2^{118}$ random numbers in total.  We have a suggestion regarding usage that may help make all the numbers more easily available.  One can on initialization jump ahead to a later portion in the Weyl sequence to use that data instead of the data at the beginning of the Weyl sequence.  For example, on the first run one might set the x, w, s as above.  On the second run after setting x, w, and s, one could multiply w by say a trillion, w *= 1000000000000; \mbox{and use that portion of the Weyl sequence.}  On the third run one could multiply w by 2 trillion, on the fourth run by 3 trillion and so on.  Using this method one could use all $ 2^{64} $ numbers available from a given Weyl sequence.\newline

\section{Parallel Usage and Jump Ahead}

This RNG is well suited to parallel processing.  Creating a separate stream of data is simply a matter of choosing a unique integer.  Assume we have a supercomputer with N cores and n identifies a given core (0 $\leq$ n $<$ N).  One could initialize the state for each instance of the RNG as shown in the example above. That is, one would set \mbox{x = w = s} using either init\_rand\_digits(n) or seed[n].

An alternative way to produce streams would be to use the jump ahead method mentioned in the previous section.  Assume that seed is an entry from seed.h. One could produce a stream of length of a trillion for core n as shown below.  The jump ahead routine is in the msws32 include file.  It advances the Weyl sequence by the specified amount.  It also randomizes the x value so that the streams produced will be relatively random.

\begin{verbatim}

#include "msws32.h"
x = w = s = seed;
if (n>0) jump_ahead(n * 1000000000000);

\end{verbatim}

Given that the length of a Weyl sequence is $ 2^{64} $, one could generate about 18 million streams of length a trillion using just one seed.  Should more streams be needed, one could use additional seeds.  This method would reduce the total number of seeds needed for a given application.

\section{64-bit Output}

In this section we describe a 64-bit output version of the RNG.  A 64-bit output could be produced by simply calling the 32-bit version twice.  We discovered however that if we take advantage of instruction-level parallelism (ILP) on modern CPUs, we could generate the 64 bits much faster.   See below. 

\begin{verbatim}

uint64_t x1 = 0, w1 = 0, s1 = 0xb5ad4eceda1ce2a9;
uint64_t x2 = 0, w2 = 0, s2 = 0x278c5a4d8419fe6b;
inline static uint64_t msws64() {
   uint64_t xx;
   x1 *= x1; xx = x1 += (w1 += s1); x1 = (x1 >> 32) | (x1 << 32);
   x2 *= x2;      x2 += (w2 += s2); x2 = (x2 >> 32) | (x2 << 32);
   return xx ^ x2;
} 
\end{verbatim}
On an Intel Core i7-9700, this version is about 40\% faster than calling the 32-bit version twice.

The jump ahead routine for the 64-bit generator is in the msws64 include file.  Assume that seed1 and seed2 are entries from seed.h.  One could produce a stream of length a trillion for core n as shown below.

\begin{verbatim}

#include "msws64.h"
x1 = w1 = s1 = seed1;
x2 = w2 = s2 = seed2;
if (n>0) jump_ahead2(n * 1000000000000);
\end{verbatim}

\section{Floating-Point Numbers}

In random number generation, a floating-point number in the range of 0 to 1 is often required.  In this section, we address the generation of floating-point numbers.

For the msws32 generator, one can simply divide by $2^{32}$ as shown below.

\begin{verbatim}

#define two32 4294967296.
double r;
r = msws32() / two32;

\end{verbatim}

This will produce a 32-bit precision floating-point number in [0,1). A good compiler will convert a division by a constant to a multiplication\footnote{For the gcc compiler, this occurs with optimization -O1 or higher and can be verified by setting -save-temps and looking at the .s output file.}.  If your compiler does not do this then you would multiply by 2.32830643653 86962890625e-10 instead of dividing by $2^{32}$. 

Similarly, for a 53-bit precision floating-point number, using msws64,

\begin{verbatim}

#define two53 9007199254740992.
double r;
r = (msws64()>>11) / two53;

\end{verbatim}

This will produce a 53-bit precision number in the range of [0,1).  If your compiler does not convert a division by a constant to a multiplication, you would multiply by 1.1102230246251565404236316680908203125e-16 instead of dividing by $2^{53}$. 
\newpage
One more option for producing 32-bit precision floating-point nuimbers would be to use msws64() to produce two 32-bit outputs as shown below.

\begin{verbatim}

#define two32 4294967296.
double r1, r2;
union {
   uint64_t i64;
   uint32_t i32[2];
} u;
u.i64 = msws64();
r1 = u.i32[0] / two32;
r2 = u.i32[1] / two32;

\end{verbatim}

\section{Statistical and Timing Results}

The msws generator was subjected to the BigCrush \cite{Lecuyera} and PractRand \cite{Dotyhumphrey} tests.  25000 random seed values were tested with no significant failures.   A square is nonlinear and for this reason the data is likely to be quite superior to data produced by any linear based RNG.   This generator passes the tests of Linear Complexity and Matrix Rank. 

The time to generate and sum one billion random numbers was computed using an Intel Core i7-9700 3.0 GHz processor running Cygwin64 with gcc version 11.2.0. To provide a basis of comparison, we also timed Marsaglia's xorwow \cite{Marsaglia} and Vigna and Blackman's xoroshiro \cite{Vigna}. 

\begin{table}[h!]
  \centering
  \caption{Time to generate and sum one billion random numbers}
  \label{tab:table2}
  \begin{tabular}{ccc}\\

    \toprule
    RNG & Precision & Time (sec)\\
    \midrule
    msws64 & 32 & 0.87\\
    xoroshiro & 53 & 1.09\\
    msws32 & 32 & 1.18\\
    xorwow & 32 & 1.26\\
    msws64 & 53 & 1.29\\
    \bottomrule
  \end{tabular}
\end{table}

The fastest case was running msws64 to produce two 32-bit precision floating-point numbers.  This was faster than xoroshiro.  Vigna suggests using the upper order bits from xoroshiro to produce floating-point numbers since the low order bits have statistical defects \cite{Vigna}.  All 64 bits from msws64 pass the statistical tests.  For this reason two 32-bit outputs can be produced.  If 32-bit precision is adequate (which is most often the case) msws64 may be the fastest RNG for producing floating-point numbers.  

\section{Cryptography}

Here we briefly discuss the subject of cryptography. The fundamental requirement for a cryptographically secure RNG is that it be difficult to determine the internal state by examining the outputs.  This RNG is nonlinear and hence the only attack would probably be brute force.  The state vector has three 64-bit elements: x, w, and s.  Using sheer brute force, one could crack the state with $(2^{64})^{3}$ or $2^{192}$ combinations of x, w, and s.  A more efficient attack which uses known available information might be as follows:  The available information we have is a sequence of known outputs and also the operations performed in the RNG (square, add Weyl sequence, and shift).  Now, let's consider three subsequent x values. This would allow us to determine two subsequent w values and then an s value.  This would provide a potential  x, w, s combination which could be verified by checking if the generated outputs matched the known outputs.  A sequence of known outputs would reduce the number of bits to examine in x from 64 to  32.  So, if we look at three x values at time, it would require $(2^{32})^{3}$ or $2^{96}$ values.  This is smaller than $2^{192}$ but is still quite a large number.  It would still be impractical to crack the state in reasonable time even with this more efficient method. We conclude that the RNG is suitable for cryptography. One could significantly increase the complexity by simply xoring the output from two different msws generators.  This would increase the complexity to  $(2^{64})^{3} * (2^{32})^{3}$ or $2^{288}$.

\section{Running the Middle Square with an RNG}

In this paper we have proposed using a Weyl sequence to run the middle square.  The Weyl sequence is very fast and produces satisfactory results. It may be of interest, however, to note that any reasonably good RNG with uniform output could be used instead of the Weyl sequence.    Assume that rand64() is a 64-bit RNG. The following shows how to create a middle-square version of this RNG:

\begin{verbatim}

uint64_t x = 0;
inline static uint32_t msrand() {
   x *= x; x += rand64(); return x = (x>>32) | (x<<32);
}

\end{verbatim}

This will produce a 32-bit output that would most likely pass the tests of Linear Complexity and Matrix Rank. 

\section{Conclusion}

The middle square was invented at the very beginning of computer \mbox{science}.  Modern 64-bit computing architecture has made it possible to create a \mbox{usable} version which has a sufficiently long period ($2^{64}$ per stream). Processing speed is comparable with the fastest RNGs. An easy to use stream capability makes this generator quite suitable for parallel processing.   A square is nonlinear and gives this generator an advantage in quality of data over linear-based generators.\newline\newline
A free software package with example programs  is available at \newline\url{http://mswsrng.wixsite.com/rand}\newline \newline
Note:  A counter-based version of this RNG is now available.  See arXiv paper "Squares: A Fast Counter-Based RNG" \cite{Widynski}.

\paragraph{Acknowledgements}
This project began in 2012 when I was reminded of an RNG I had created many years ago.  I ran some Google searches on random number generation and by chance came to the Wikipedia article on the Mersenne Twister \cite{Wiki1}\cite{Matsumoto}.  Mersenne Twister is the most widely used RNG.  The article stated that it passed most but not all of the tests in TestU01 \cite{Lecuyera}.  I thought I would see how my RNG would do with TestU01.  I downloaded TestU01 and Cygwin on my home computer and tried my RNG.  It passed SmallCrush but not Crush or BigCrush.  I ran tests on revised RNGs to see if I could pass BigCrush. This became sort of a hobby.  I eventually ended up with a middle-square RNG based on floating-point arithmetic.  This RNG did pass BigCrush.  I decided to present it to the ACM (Transactions on Mathematical Software).  One of the reviewers objected to the floating-point arithmetic.  It would not be compatible across different computing platforms.  I happened to locate the Wikipedia article on \lq\lq Integer Overflow\rq\rq \cite{Wiki2}.  I had previously tried integers but thought I had to store the entire result when squaring.  The Wikipedia article stated that C used modulo arithmetic on overflow.  I had a quick moment of insight when reading that article that I could use only half of the square.  I went home and programmed this and the RNG passed BigCrush on the first attempt.  I presented this new RNG to the ACM (Transactions on Modeling and Computer Simulation) and this time the reviewer had a question about uniformity.  I did extended analysis on this and discovered that middle squaring as I had implemented it was not quite uniform.  Some numbers came up more frequently than others.  I considered this problem and tried an additive solution which worked reasonably well.  However, I emailed Richard Brent, creator of the xorgens RNG \cite{Brent}.  He suggested that I use the Weyl sequence to obtain a uniform output.  This merely involved rearranging the order of operations in the RNG.  The Weyl sequence was added after squaring instead of before squaring.  This surprisingly simple solution solved the problem.  The distribution produced is quite comparable with the Mersenne Twister distribution which is known to be a uniform distribution. I am thankful for the anonymous ACM reviewers' inputs and particularly for Brent's suggestion regarding uniformity. These led to the RNG that is here published.  

I would like to thank the creators of the BigCrush \cite{Lecuyera} test without which this would not have been possible.  Pierre L'Ecuyer and Richard Simard answered emails I sent regarding some of my early RNGs.  Their responses were helpful and led to improvements. 

I would like to thank Sebastiano Vigna and David Blackman for publishing \lq\lq xoroshiro128+\rq\rq \cite{Vigna}.  It was from this program that I learned of the C construct that produces a rotate instruction which helps make the RNG so efficient.  I would like to thank George Marsaglia for publishing \lq\lq Xorshift RNGs\rq\rq\cite{Marsaglia} where I initially learned of the Weyl sequence idea. I would like to thank  Mutsuo Saito and Makoto Matsumoto for publishing \lq\lq XSadd\rq\rq \cite{Saito} where I learned of ``inline static'' which improves efficiency.

I would like to thank Chris Doty-Humphrey the creator of PractRand \cite{Dotyhumphrey}. PractRand provided an additional set of statistical tests for the generator.  I would like to thank Paul Doe for providing the PractRand statistics for sparse seed values.

I would like to thank an Indian friend who recommended the different digits idea for initialization. This produces change in each 4-bit section of the Weyl sequence on each iteration of the RNG.

I would like to thank D. E. ShawResearch for publishing ``Parallel Random Numbers: As Easy as 1, 2, 3'' \cite{Salmon}.  This paper made me aware of the requirements for parallel random numbers.

I would like to thank Donald Knuth for a suggestion regarding an early version of the RNG. Even though gcc did compile the program correctly he felt other compilers might not.  The program was revised to account for this. 

I am thankful for the freely available computing and research tools that made this project possible:  Google, Wikipedia, and Cygwin.  WiX and Microsoft provided free web and cloud services.  I should also mention arXiv which provides a freely available publication. 

The initial article did not consider cryptography.  It was presented to the Data Structures and Algorithms section of arXiv.  The arXiv moderators decided to publish it in the Cryptography and Security section.  It is for this reason that the subject of cryptography is addressed in this version.  I agree with the arXiv moderators that the RNG is suitable for cryptography and am thankful that they pointed this out.

I am thankful for the creation of the 64-bit computer.  Middle squaring works best with large squares.  John von Neumann suggested  \lq\lq more digits than ten\rq\rq \cite{Neumann}.  64 bits is equivalent to 16 hexadecimal digits.

I am thankful for the C programming language which provides a simple compact implementation.

I would like to thank Hermann Weyl for his work on equidistribution which laid the foundation for uniformity in this generator \cite{Weyl}. The reader is referred to the Wikipedia \lq\lq Weyl Sequence\rq\rq  article \cite{Wiki3} which shows a nice parallel between irrational sequences and integer sequences.

I would like to thank my high school computer science teacher Mr. Knight.  It was in his class that I initially learned of middle square.

I would like to thank Stanislaw Ulam the inventor of the Monte Carlo method \cite{MetropolisUlam}.  Random number generation on computers began with Monte Carlo problems. 

I would like to thank John von Neumann for creating the middle-square method.  His original idea for random number generation turns out to be one of the best methods.    

I would like to thank Barbara and her family for their support throughout this effort.

\newpage

\ifx\printbibliography\undefined
    \bibliographystyle{plain}
    \bibliography{msws}

@ARTICLE{Marsaglia,
   AUTHOR="George Marsaglia",
   TITLE="Xorshift RNGs",
   JOURNAL="Journal of Statistical Software",
   YEAR=2003,
   VOLUME=8,
   NUMBER=14,
   URL="www.jstatsoft.org/v08/i14/paper"
}

@ARTICLE{Neumann,
  AUTHOR="John von Neumann",
  TITLE="Various Techniques Used in Connection with Random Digits",
  JOURNAL="A.S. Householder, G.E. Forsythe, and H.H. Germond, eds., Monte Carlo Method, National Bureau of Standards Applied Mathematics Series",
  YEAR=1951,
  VOLUME=12,
  PAGES={36-38},
}

@ARTICLE{Lecuyera,
  AUTHOR="P. L'Ecuyer and R. Simard",
  TITLE="TestU01: A C Library for Empirical Testing of Random Number Generators",
  JOURNAL="ACM Transactions on Mathematical Software",
   YEAR=2007,
  VOLUME=33,
  URL="http://simul.iro.umontreal.ca/testu01/tu01.html",
}

@ARTICLE{Matsumoto,
  AUTHOR="M. Matsumoto and T. Nishimura",
  TITLE="Mersenne Twister: A 623-Dimensionally Equidistributed Uniform Pseudo-Random Number Generator",
  JOURNAL="ACM Transactions on Modeling and Computer Simulation",
  YEAR=1998,
  VOLUME=8,
  NUMBER=1,
  PAGES={3-30},
  URL="http://www.math.sci.hiroshima-u.ac.jp/~m-mat/MT/emt.html",
}

@ARTICLE{Brent,
 AUTHOR="R. P. Brent",
 TITLE="Some Long-Period Random Number Generators using Shifts and Xors",
 YEAR=2006,
 JOURNAL="ANZIAM Journal",
 VOLUME=48,
 PAGES="C188-C202",
 URL="http://maths-people.anu.edu.au/~brent/pub/pub224.html",
}

@MISC{Dotyhumphrey,
 AUTHOR="C. Doty-Humphrey",
 TITLE="Practically Random: C++ library of statistical tests for RNGs",
 YEAR=2018,
 NOTE="version 0.94",
 URL="http://pracrand.sourceforge.net",
}

@MISC{Vigna,
 AUTHOR="S. Vigna and D. Blackman",
 TITLE="xoroshiro128+",
 YEAR=2018,
 URL="http://xoshiro.di.unimi.it/xoroshiro128plus.c",
}

@MISC{Saito,
 AUTHOR="M. Saito and M. Matsumoto",
 TITLE="XSadd",
 YEAR=2014,
 URL="http://www.math.sci.hiroshima-u.ac.jp/~m-mat/MT/XSADD/",
}

@CONFERENCE{Salmon,
   AUTHOR="J.K. Salmon and M.A. Moraes and R.O. Dror and D.E. Shaw",
   TITLE="Parallel Random Numbers: As Easy As 1, 2, 3",
   BOOKTITLE="Proceedings of the 2011 International Conference for High Performance Computing, Networking, Storage and Analysis, ACM, New York",
   YEAR=2011,
   PAGES={16:1-16:12},
}

@ARTICLE{Wiki1,
 AUTHOR="Wikipedia Contributors",
 TITLE="Mersenne Twister",
 YEAR=2018,
 JOURNAL="Wikipedia, The Free Encyclopedia",
 URL="https://en.wikipedia.org/w/index.php?title=Mersenne_Twister&oldid=864897141",
}

@ARTICLE{Wiki2,
 AUTHOR="Wikipedia Contributors",
 TITLE="Integer Overflow",
 YEAR=2018,
 JOURNAL="Wikipedia, The Free Encyclopedia",
 URL="https://en.wikipedia.org/w/index.php?title=Integer_overflow&oldid=874320318",
}

@ARTICLE{Wiki3,
 AUTHOR="Wikipedia Contributors",
 TITLE="Weyl Sequence",
 YEAR=2018,
 JOURNAL="Wikipedia, The Free Encyclopedia",
 URL="https://en.wikipedia.org/w/index.php?title=Weyl_sequence&oldid=878733044",
}

@ARTICLE{Weyl,
   AUTHOR="H. Weyl",
   TITLE="Ueber die Gleichverteilung von Zahlen mod. Eins",
   YEAR=1916,
   JOURNAL="Math. Ann.",
   VOLUME=77,
   NUMBER=3,
   PAGES={313-352},
}

@ARTICLE{Widynski,
  AUTHOR="Bernard Widynski",
  TITLE="Squares: A Fast Counter-Based RNG",
  JOURNAL="arXiv:2004.06278",
  YEAR=2020,
  URL="http://arxiv.org/abs/2004.06278"
}

@ARTICLE{MetropolisUlam,
   AUTHOR="N. Metropolis and S. Ulam",
   TITLE="The Monte Carlo Method",
   YEAR=1949,
   JOURNAL="Journal of the American Statistical Association",
   VOLUME=44,
   NUMBER=247,
   PAGES={335-341},
}
\else\printbibliography\fi

%\appendix
%\section*{Appendix}
%\setcounter{section}{0}
%\section{Usage Notes}

\end{document}